# Inception Architecture and Residual Connections in Classification of Breast Cancer Histology Images

Mohammad Ibrahim Sarker[1], Hyongsuk Kim[1,] Denis Tarasov[2], Dinar Akhmetzanov [2]

[1] Chonbuk National University, Jeonju-si, jeollabuk-do, 54896 Republic of Korea
[2] Meanotek AI Research, Sibirsky Trakt 34 building 4-309, Kazan, Russian Federation
sarkeribrahim@gmail.com , dtarasov@meanotek.io

**Abstract.** This paper presents results of applying Inception v4 deep convolutional neural network to ICIAR-2018 Breast Cancer Classification Grand Challenge, part a. The Challenge task is to classify breast cancer biopsy results, presented in form of hematoxylin and eosin stained images. Breast cancer classification is of primary interest to the medical practitioners and thus binary classification of breast cancer images have been under investigation by many researchers, but multi-class categorization of histology breast images have been challenging due to the subtle differences among the categories.
In this work extensive data augmentation is conducted to reduce overfitting and effectiveness of committee of several Inception v4 networks is studied. We report 89% accuracy on 4 class classification task and 93.7% on carcinoma/non-carcinoma two class classification task using our test set of 80 images.

## 1   Introduction

Cancer is identified as the second highest cause for mortality in the developing countries [3]. Among the various kinds of cancers, women are highly prone to get affected by breast cancer and ovarian cancer. Breast cancer denotes a state in which the cancerous cells are developed in the breast tissues. It results in very high mortality among women.  Once the tumor is suspected based on mammography  or ultrasound study, biopsy is recommended by the oncologists to verify diagnosis and obtain detailed information on tumor type and staging. The biopsy of the breast can reveal various conditions. Dataset of ICIAR-2018 grand challenge is annotated with four classes of possible histological findings: normal (no finding), benign, and two types of malignant tumor types: ductal cell carcinoma *in situ* (DCIS) and invasive carcinoma, which are two most common histological types of breast cancer. The in situ refers to a condition where the cancerous cells are present only within the mamalian gland ductile tubular system and do not spread to other places whereas the invasive category poses the risk of getting spread to other locations. Manual investigation of these images by pathologists is time consuming, error-prone and subject to inter-variability among the experts. Hence automated solutions are sought for this purpose.

Among the various computational techniques, image processing and machine learning techniques have been extensively adopted to diagnose breast cancer. In the recent times, deep learning techniques have established its potential in computer assisted medical image analysis.

Our work is built upon and extends recently published approach of patch-wise classification using Convolutional Neural Networks (CNNs) [2]. We attempt to extend the approach presented in [2] in three ways:

1. Very deep Inception v4 [9] architecture has been utilized instead of relatively simple network

2. Extensive dataset augmentation with a number of techniques, specifically designed for histology slides, has been conducted to prevent overfitting

3. A committee of neural networks formed by combining multiple checkpoints from same training run is utilized

The main findings from this work include:

1. Inception v4 gives good results on provided dataset even with basic data augmentation presented in [2]

2. Additional data augmentation using the proposed method can significantly improve results

3. With data augmentation, specific techniques for stained slides normalization is not necessary to obtain good classification accuracy. This is in contrast to the results reported in [2]

4. Combining multiple checkpoints from the same training run provides improvements in accuracy for the task in hand.

The remaining of the paper is structured as follows: Section 2 presents the related work in detecting breast cancer. Section 3 describes the proposed methodology. Section 4 discusses the results and Section 5 concludes the paper along with few insights on the future directions.

## 2   Related Work

There are many methods to carry out breast cancer classification such as DNA-based [3] classification and computer vision [2] based to apply validated algorithms and present visually with graphics. Here we focus on the method based on computer vision. We can divide computer vision proposes into parts: traditional machine learning way and recurrent popular way.

For the traditional machine learning way, researchers typically use hand-crafted algorithms: Linear Regression, Nearest Neighbor search, Multilayer Perceptron, Softmax Regression and Support Vector Machines (SVN) [1]. Most of these methods are based on hand-crafted features, which hinder the classification accuracy and robustness.

Initial research focused on nuclei analysis in the view of classifying malignant and benign tumors. Kowal et al. [13] has adopted various clustering procedures for delineating the nuclei. The methodology has been evaluated on fine needle biopsy microscopic images. Various kinds of features such as morphological, topological and texture related features have been extracted from the segmented nuclei and utilized to train the classifier. Accuracies ranging between 84% and 93% have been achieved on 500 images acquired from 50 patients.

Similarly, Filipczuk et al. [14] and George et al. [15] have elicited nuclei-based features from fine needle biopsies. Initially, circular Hough transform is employed for selecting the candidates for nuclei. Subsequently, the machine learning techniques on shape and texture features of the candidate nuclei have been adopted to eliminate the false positive nuclei candidates. George et al. [15] has refined the nuclei segmentation process through watershed segmentation. Filipczuk et al. [14] have been able to achieve an accuracy of 98.51% through majority voting over 11 images for each of the 67 patients. On the other hand, George et al. [15] have obtained an accuracy ranging between 71.9% and 97.15% while classifying 92 individual images.

In addition to nuclei-based characteristics, tissue organization has also been taken into account for the binary classification of more intricate images [16]. Accuracies ranging from 70% to 100% are reported when evaluating on 70 images obtained from a private 40× magnification breast histology H&E dataset.

Some other research works have placed attention in classifying the breast cancer histology images into multi-class (three class) categorization. To mention a few, Brook et al. [17] and Zhang et al. [17] have attempted to classify the breast cancer tissue images into normal, in situ carcinoma and invasive carcinoma classes.

Subsequently, a cascade classification methodology has been advocated for breast cancer classification [17]. The images are initially subjected to subsets of curvelet transformation. Then, local binary pattern (LBP) features are extracted and randomly provided as input to the first set of parallel SVM classifiers. This system has been able to accomplish an accuracy of 97% accuracy with 0.8% rejection rate.

Recently, due to the availability of powerful GPU implementation and large training dataset, deep learning based methods have achieved great success in computer vision and hence were applied to breast cancer histology classification task. Han et al [5] proposed class structure-based deep convolution neural network (CSDCNN), which is the first to leverage hierarchical feature representation. Wang et al [12] presented a deep learning-based system for breast cancer classification with slide-based model.

On 4-class classification problem, discussed in this paper, little previous work exist. The problem was first formulated in [2] which also presented a solution using custom convolutional neural network and SVM classifier. Authors reported 86% accuracy on 4 class classification and 90% accuracy on 2 class classification task.

### 2.1 Datasets

Dataset, provided by ICIAR-2018 Grand Challenge contains 400 images (2040 × 1536 pixels) classified into 4 classes: normal, benign, carcinoma in situ and invasive carcinoma. Dataset is balanced comprising of 100 images for each class. Since at the time of this writing test dataset was not available to us, the results are reported for a subset of 80 images, randomly chosen from the training set. Thus, the training set comprises of 320 images and test data is composed of 80 images.

## 3    Methods and algorithms

### 3.1. Datasets

Dataset, provided by ICIAR-2018 Grand Challenge contains 400 images (2040 × 1536 pixels) classified into 4 classes: normal, benign, carcinoma in situ and invasive carcinoma. Dataset is balanced comprising of 100 images for each class. Since at the time of this writing test dataset was not available to us, the results are reported for a subset of 80 images, randomly chosen from the training set. Thus, the training set comprises of 320 images and test data is composed of 80 images.

### 3.2. Preprocessing

#### 3.2.1. Basic Preprocessing and Data Augmentation

The data augmentation procedure reported in following [2] is employed in this work. We divide original image is divided into on 12 contiguous non-overlapping patches. For each patch, we generated eight different patches are generated by combining k·π/2 rotations, with k = {0, 1, 2, 3}, and vertical reflections. Unlike previous works, we did not apply staining normalization is not applied, since one of objectives of this work is was to develop a method that can work without it.

#### 3.2.2. Extended data augmentation

For each new training batch, we apply following transformations are applied:
   *Elastic deformations along x and y dimensions*
  We generate stretches and compressions of all images are generated with scaling factor ranging from 0.7 to 1.3. Scaling factor is chosen randomly from uniform distribution for each image sample.
   *Brightness and contrast variations*
   Brightness and contrast changes are commonly defined using $g(x)=\alpha f(x)+\beta$, equation where $\alpha$ is considered to be the relative brightness and $\beta$ is considered to be the contrast. Both values are chosen randomly from uniform distribution in the range [-0.2:0.2]
   *Gaussian blur*
   Gaussian blur with value of $\sigma$ chosen randomly from range [0.3:0.6] is was added to the images by applying convolutional filter.
   *Uniform noise*
   Uniform noise was added by randomly changing 1% of all pixels in the image to color values randomly chosen from other image parts.

*Resampling*

Images are were randomly resized to 80-99% of their original size and scaled back to original size using bicubic anti-aliasing.

### 3.1 Neural Network Architecture and Training

All neural networks were implemented using Tensorflow (www.tensorflow.org) library. We tried both version 1.2 and 1.4 version of the library and obtained slightly different results. Results in the paper are provided for 1.4. unless otherwise is noted.

*Baseline Simple CNN*

Since no reference results were published for this dataset prior to completion, we implement simple convolutional network to serve as baseline. Baseline CNN apply 2x2 convolution to image patches scaled to 256x256, followed by 2x2 max-pooling, 2x2 convolution, 4x4 convolution, 4x4 max-pooling, and fully connected layer with 512 units. Output layer was simple softmax over 4 classes.

*Inception CNN*

The Inception deep convolutional architecture has been introduced as GoogLeNet in [10]. It is now commonly called "Inception-v1". This model has demonstrated superior results in related histological image classification challenge [12]. Next generations of Inception added batch normalization (v2) and residual blocks demonstrated to facilitate training of very deep networks earlier [6]. Detailed description of Inception v4 architecture, used in this paper is given in [9]. We use implementation available from Tensorflow repository with topmost layer changed into 4-class softmax. Weights of other layers were initialized from checkpoint pre-trained on ImageNet dataset.

### 3.2 Predictions

Predictions for whole images are done using majority vote approach that was shown to be best when aggregating predictions from multiple patches [2]. For each test image 12 patches are generated and predictions are obtained for each of them, then most frequent prediction is chosen as final. When using committee of networks, each network is used to predict class of each image patch, and majority vote is used to select final answer, in similar way.

## 4 Results and Discussion

### 4.1 Results on Patchwise Classification

Accuracy on patchwise classification task is presented in table 1. Unfortunately, we lack reference results for comparison. In [2], patchwise classification accuracy on 4 classes is reported to be 72.5 for their original CNN model and 72.9 for using SVN classifier with features extracted from the topmost CNN layer. The results obtained from the proposed method are substantially better. However, these results have to be interpreted with extreme caution, since ICIAR-2018 dataset is larger (we used 320 images for training, while [2] use 75% of 253 training images), our test sets are different and official competition test set is not available to us at the time of this writing. In [2] two different test sets have been used. They are "Initial" (20 images) and "Extended" (16 images) described as "images of increased ambiguity". The test set of this work contains 80 randomly sampled images. We compare our results to accuracy reported in [2] for "Initial" (simpler) test set.

**Table 1.** Accuracy on patchwise classification task.

| Model | Accuracy on training set | Accuracy on test set | Training time on 1080 GTX, days |
|---|---|---|---|
| Simple CNN | 57.0% | 59.0% | 2 |
| Inception v4 | 86.0% | 75% | 5 |
| Inception v4 + data augmentation | 83.0% | **78.5%** | 8 |

### 4.2 Accuracy on Whole Image Classification Task

Accuracy on whole image classification task is presented in table 2. In [2] best resutls are 80% for CNN and 85% for CNN+SVN. Predictably, our results also exceed these published in [2].

**Table 2.** Accuracy on Whole Image Classification Task.

| Model | Accuracy on test set |
|---|---|
| Simple CNN | 70.0% |
| Inception v4 | 82% |
| Inception v4 + dataset augmentation | 87.5% |
| Committee of 3 Inception v4 networks | **89.5%** |

**Table 3.** Accuracy on Individual Classes.

| Model | Normal | Benign | InSitu | Invasive |
|---|---|---|---|---|
| Simple CNN | 65% | 53% | 51% | 64% |
| Inception v4 | 80% | 60% | 61% | 77% |
| Inception v4 + dataset augmentation | **83%** | **66%** | **64%** | **83%** |

On two class classification task carcinoma/non-carcinoma our model achieved 93.7% accuracy, improving on 90% reported in [2].

### 4.3 Visualization of feature detectors

To better understand overfitting issues, we visualized feature detectors of the first layer of trained Inception network. There are mostly edge and central activation detectors, tuned to nuclei and cell features. Apparently many of them are not used (all weights are the same) and also we can see redundant almost identical detectors. Our future work will explore possibility of removing unnecessary feature detectors from Inception architecture, to better tune it for breast cancer histology classification.

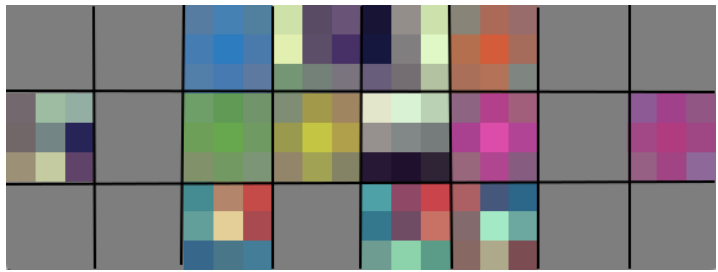

Figure 1. Visualization of sample feature detectors in the first layer of trained Inception network. RGB channels are combined into single image to facilitate analysis.

## 5 Conclusions

We proposed an approach for the classification of breast cancer histology images based on convolution neural networks (CNNs). We demonstrate that Inception-v4 dramatically improves on simple CNN baseline. Our experiments suggest that it also outperforms custom CNN architectures that was used in previous studies for 4 class classification, when extensive data augmentation strategy is applied. Interestingly, good results (89% accuracy on 4 class classification task) can be obtained without using specialized staining image normalization methods, contrary to all previous results on similar datasets.